\documentclass[11pt,twoside]{article}

\usepackage{asp2014}
\usepackage{xcolor}

\aspSuppressVolSlug
\resetcounters

\begin{document}

\title{In-Situ High Performance Visualization for Astronomy \& Cosmology}

\author{Nicola~Tuccari,$^{1,2}$ Eva~Sciacca,$^1$ Yolanda Becerra$^3$, Enric Sosa Cintero$^3$, Robert Wissing$^4$, Sijing Shen$^4$, Emiliano Tramontana$^2$
}
\affil{$^1$INAF Astrophysical Observatory of Catania, Catania, Italy; \email{nicola.tuccari@inaf.it}}
\affil{$^2$Università di Catania, Dipartimento di Matematica e Informatica, Catania, Italy}
\affil{$^3$Barcelona Supercomputing Center, Barcelona, Spain}
\affil{$^4$Institute of Theoretical Astrophysics, University of Oslo, Oslo, Norway}

\paperauthor{Eva~Sciacca}{eva.sciacca@inaf.it}{0000-0002-5574-2787}{INAF}{Astrophysical Observatory of Catania}{Catania}{CT}{95123}{Italy}
\paperauthor{Nicola~Tuccari}{nicola.tuccari@inaf.it }{0009-0004-7802-2602}{Università di Catania}{Dipartimento di Matematica e Informatica}{Catania}{CT}{95123}{Italy}



\begin{abstract}
The Astronomy \& Cosmology (A\&C) community is presently witnessing an unprecedented growth in the quality and quantity of data coming from simulations and observations. Writing results of numerical simulations to disk files has long been a bottleneck in high-performance computing. To access effectively and extract the scientific content of such large-scale data sets appropriate tools and techniques are needed. This is especially true for visualization tools, where petascale data size problems cannot be visualized without some data filtering, which reduces either the resolution or the amount of data volume managed by the visualization tool.

A solution to this problem is to run the analysis and visualization concurrently (in-situ) with the simulation and bypass the storage of the full results. In particular we use Hecuba, a framework offering a highly distributed database to stream A\&C simulation data for on-line visualization. We will demonstrate the Hecuba platform integration with the Changa high performant cosmological simulator and the in-situ visualization of its N-body results with the ParaView and VisIVO tools.
\end{abstract}



\section{Introduction}

Astrophysical observations and simulation codes executed on high-performance supercomputers generate massive data volumes, often reaching petabyte scales. These enormous datasets present significant challenges in terms of storage, access, and analysis, which are crucial for facilitating scientific discoveries \cite{hey2009the}. Pre-exascale systems provide remarkable opportunities to scale HPC applications in Astrophysics and Cosmology (A\&C), particularly for the high-performance visualization needed to interpret their results.

The dominant paradigm for scientific visualization over the years has been post-processing. This approach involves saving the data generated by simulation codes to permanent storage, so visualization and analysis tools need to load this data after it is stored. An emerging processing paradigm is in-situ visualization. It involves visualizing and analyzing data as it is generated, allowing the user to create graphical output while the simulation executes.

Only HPC resources have the capability to process such vast amounts of data. However, the primary source of performance and scalability challenges lies in parallel file systems. Additionally, working with files imposes rigid application workflows that require synchronization and complex code, making it difficult to adapt to new requirements or hardware changes.

Key-Value (KV) databases are widely used in data analytics and represent
a realistic alternative. They are well-suited for scientific applications such as time-series or spatial data. Furthermore, they allow analyzing partial results and react, for instance, by discarding a cosmological simulation as soon as a certain event occurs.

In this work we exploit Hecuba that allows for a particular type of in-situ visualization: in-transit visualization \cite{moreland2011examples}, i.e. analysis and visualization is run on I/O nodes that receive the full simulation results but write information from analysis or provide run-time visualization. We focus on the Hecuba platform integration with the Changa high performant cosmological simulator and the in-situ visualization of its N-body results with the ParaView and VisIVO tools.

\section{Data, Tools and Methodology}

\subsection{ChaNGa Cosmological Data}

ChaNGa\footnote{\url{https://github.com/N-BodyShop/changa}} (Charm N-body GrAvity solver) is a code that performs collisionless N-Body simulations. It supports cosmological simulations with periodic boundary conditions in comoving coordinates, as well as simulations of isolated stellar systems. Additionally, it can incorporate hydrodynamics using the Smooth Particle Hydrodynamics (SPH) method. Gravity calculations are performed using a Barnes-Hut tree algorithm, which includes hexadecapole expansion of nodes and Ewald summation for handling periodic forces. The code employs a leapfrog integrator with individual timesteps assigned to each particle for time integration. ChaNGa stores data using the Tipsy data format.

\subsection{Hecuba}

Hecuba\footnote{\url{https://github.com/bsc-dd/hecuba}} is a set of tools and interfaces that aims to facilitate the management and utilization of persistent data for Big Data applications. Hecuba implements an Object Mapper for Apache Cassandra \cite{CASS10} (a recognized NoSQL database) and thus allows programmers to use a common interface to access data as regular in-memory objects, regardless they are persistent (stored in disk) or they are in-memory data. Currently Hecuba implements an interface for Python and C++ programming languages. In our current work we are extending Hecuba to also support a lambda architecture \cite{MARZ15} which is a data-processing architecture defined to speed up online analysis for Big Data applications, without losing the ability to persist the output data of the applications. Thus, Hecuba will be able to support both off-line and on-line data analysis. Notice that the programmer can use the same interface to access the data whether it is in memory, in the storage or it is coming through a stream: only needs to modify the class definition of the object to indicate which kind of object it is.

\begin{figure}
    \centering
    \includegraphics[width=0.5\linewidth]{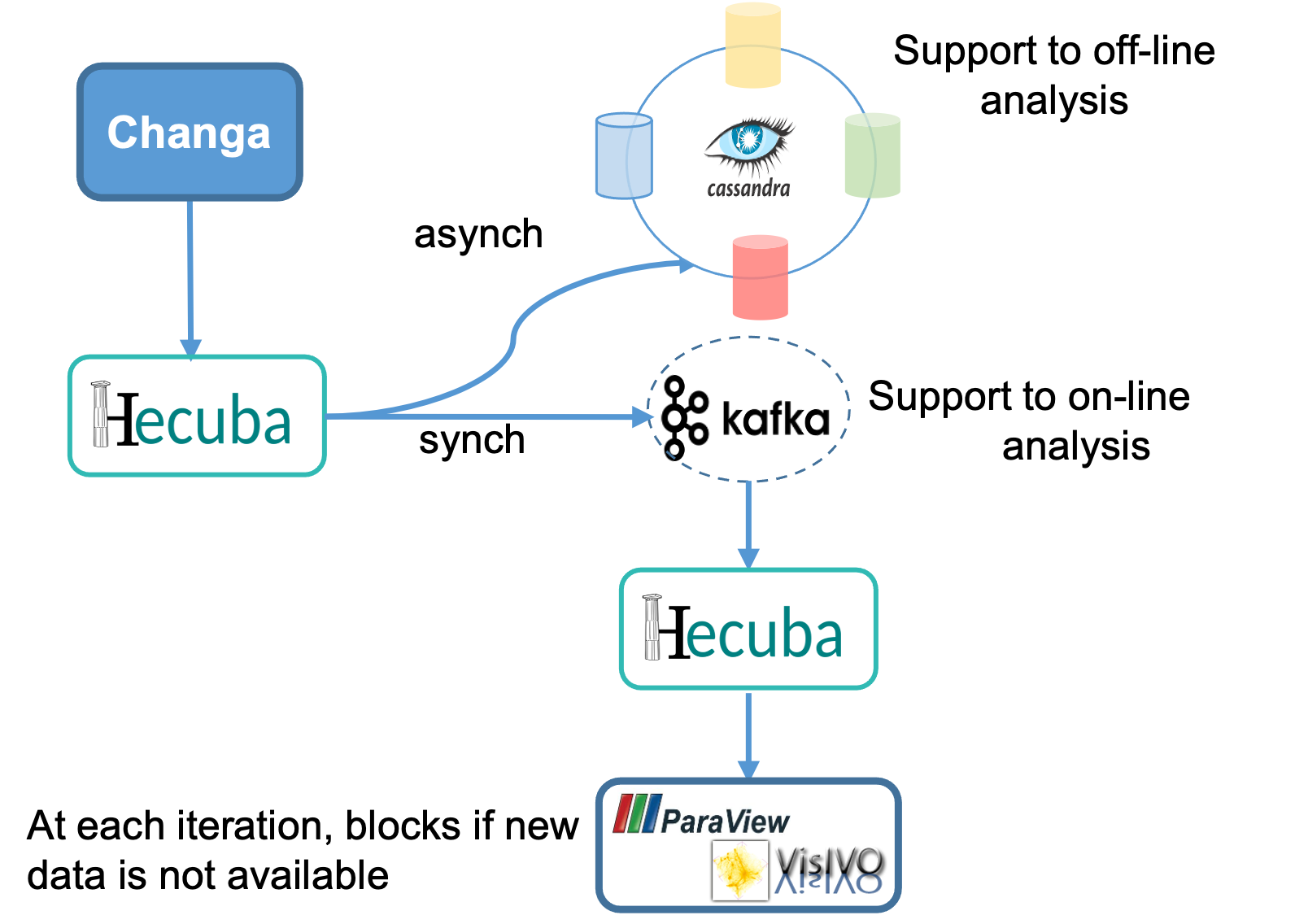}
    \caption{Hecuba architecture implementing an Object Mapper for Apache Cassandra support both off-line and on-line ChaNGa data analysis.}
    \label{fig:enter-label}
\end{figure}

\subsection{Paraview}

Paraview\footnote{\url{https://www.paraview.org/}} \cite{PVIEW05} is a tool designed to support the visualization and analysis of large scientific data sets. To this end,  Paraview supports, for example,  hardware-accelerated rendering or parallel rendering on shared and distributed memory machines. The decoupled architecture of Paraview allows the use of remote execution of a Paraview server to execute the data acquisition together with the data generating application (saving in this way moving a large amount of information between machines) and the data visualization in a desktop machine (with a more limited amount of memory and disk). Another useful capability of Paraview is its ability to support user plugins to customize the data reading and visualization process. In this work, we have implemented a custom Paraview plugin that uses the Hecuba interface to implement online acquisition of the data to visualize.

\subsection{VisIVO}
VisIVO is an integrated suite of tools and services specifically designed for the Virtual Observatory and for creating customized views of 3D renderings from astrophysical datasets \cite{sciacca2015integrated}. VisIVO Server\footnote{\url{https://github.com/VisIVOLab/VisIVOServer}} consists of three core components: VisIVO Importer, VisIVO Filter, and VisIVO Viewer. We are currently focusing on evolving VisIVO and enabling it to take advantage of new HPC facilities. 
In this work we explored the possibilities given by integrating Hecuba with VisIVO Server. By interacting with a highly distributed database, VisIVO would gain the capability to perform in-situ visualization.

\section{Preliminary Results and Future Works}
The current preliminary results for VisIVO include the development of a prototype featuring a novel Importer that retrieves simulated data from a distributed database using Hecuba's API. This prototype will serve as the basis for implementing a complete pipeline that utilizes all VisIVO Server modules to generate visualizations from the data retrieved by the reader and passed through one of VisIVO's filters.

Regarding Paraview, we have implemented a Python plugin that uses Hecuba to receive and visualize data at the same time as the simulator generates it. The goal is for the simulator to be able to run on the same infrastructure as the plugin or on a different one. Hecuba's interface allows data to be visualized as it is received without user intervention. However, we have also included the ability for the user to move backward and return to the current moment using ParaView's navigation buttons. To test the plugin, we implemented a small code that acts as a data generator. The generated data is the result of a previous Changa simulation, stored in Tipsy format files. Our producer simply reads these files and uses Hecuba's interface to send them to our ParaView plugin. We are currently in the performance analysis phase, using different data file sizes. The next step will be to integrate Changa with Hecuba so that it becomes the real producer in our experiments.

\acknowledgements This work is funded by the European High Performance Computing Joint Undertaking (JU) and Belgium, Czech Republic, France, Germany, Greece, Italy, Norway, and Spain under grant agreement No 101093441 and it is supported by the spoke "FutureHPC \& BigData” of the ICSC – Centro Nazionale di Ricerca in High Performance Computing, Big Data and Quantum Computing – and hosting entity, funded by European Union – NextGenerationEU.

\bibliography{biblio}  


\end{document}